**Instrumented mouthguards in elite sports: Validity and head acceleration event (HAE) incidence in NCAA American Football**

Mario Pasquale Rotundo[1]

Nicholas G Murray[2]

David Allan[1]

Gregory Tierney[1]

[1] Nanotechnology and Integrated Bioengineering Centre (NIBEC), School of Engineering, Ulster University, Belfast, United Kingdom

[2] School of Public Health, University of Nevada, Reno, United States

**Abstract**

*Objectives:* To determine the on-field validity of instrumented mouthguards (iMGs) in American football and to quantify head acceleration event (HAE) incidence in NCAA football players.

*Methods:* Instrumented mouthguards were fitted to 35 male NCAA football players. Head kinematic data were collected during 64 player matches. On-field validity was determined through video review with positive predictive value (PPV) and sensitivity values calculated. HAE incidence was calculated as the number of HAEs per player match and stratified by Offense and Defense positions.

*Results:* On-field validity of the Prevent Biometrics iMG in NCAA American Football indicates a sensitivity was 0.89 and PPV ranging from 0.76-0.98 based on false positive definitions. The incidence of PLA and PAA HAEs above a range of thresholds in Defense and Offense appear similar. The incidence of HAEs above 10 g was 11.2 and 11.3 HAEs per player match for Defense and Offense, respectively, while PAA incidence above 1.0 krad/s$^2$ was 5.5 and 6.9 HAEs per player match for Defense and Offense, respectively. Incidence of HAEs above 30 g was 1.6 and 2.6 per player match and 0.9 and 1.4 for HAEs above 2.0 krad/s$^2$ for Defense and Offense, respectively.

*Conclusion:* The Prevent Biometrics iMG appears suitable for measuring HAEs in elite American football. The study provides a benchmark assessment of HAE incidence in elite American Football and lays a foundation for the development of position-specific interventions aimed at reducing HAE exposure.

**Introduction**

American football is a high-impact contact sport characterised by frequent collisions during tackles, runs, blocks, and other physical confrontations. [1,2] As a result, players are at an increased risk of head injuries, including concussions.[3] Several studies in sport have explored risk factors for concussion[4,5] and various rule changes have been implemented in elite American Football to mitigate concussion risk and improve diagnosis and recovery protocols.[6] However, a growing body of evidence suggests that repetitive HAEs, even in the absence of diagnosed concussion, could have implications for long-term neurological function.[7–9] This raises significant concerns about the long-term health implications for American football players, particularly as the sport is known for the high number of HAEs that players are exposed to during games and practices.[10] In American Football, it has been demonstrated that direct helmet contact events can be of a similar magnitude to inertial head loading, suggesting that players are experiencing meaningful HAEs during gameplay that are not easily observed.[10] One feature of American football is the unique nature of each player position on the field. Measuring the incidence of HAEs is an important step to understanding the exposure of players across various positions, which may inform tailored risk-mitigation strategies to protect player brain health.

In the past, in-vivo head kinematic data has been collected using helmet-based or skin sensors which are prone to artifacts from sensor movement relative to the head.[11] However, the development of instrumented mouthguards (iMGs) has allowed for more accurate and reliable measurement of both linear and rotational head kinematics, with iMGs offering rigid coupling to the skull via the upper dentition, producing valid and reliable measurement of linear and rotational head kinematics.[11–13] The application of iMGs in sport provides a unique opportunity to measure HAEs during on-field match play and has previously been undertaken in sports such as American football, rugby, and soccer.[14–17] The Prevent Biometrics iMG (Minneapolis, MN, USA) has been previously validated in laboratory for helmeted and unhelmeted head impacts as well as on-field in the un-helmeted sport of rugby. [12,13,18] Jones et al.[12]

undertook a large-scale validation and feasibility study of multiple iMG systems and indicated that the Prevent Biometrics iMG scored highest for player fit, function and comfort, laboratory-based impact testing using a crash test dummy headform (concordance correlation coefficient value of 0.98) and on-field in rugby with a positive predictive value (PPV) of 0.94 for detecting on-field HAE. Furthermore, Tooby et al.[19] found that applying a combined iMG recording threshold of 5g (at head centre of gravity) and 400 rad/s$^2$ improved PPV to 0.99 (95% CI: 0.97-1.00), yielding a sensitivity value of 0.86 (95% CI: 0.84-0.89) for direct head impacts. The FRI-Biocore iMG system evaluated by Jones et al.[12] achieved a PPV of 0.98 and sensitivity of 0.82 in an American football-based validation study.[15]

This study aims to evaluate the on-field validity of the Prevent Biometrics iMG in NCAA Division I American football and quantify the incidence of HAEs across Defense and Offense player positions. The findings provide a benchmark assessment of HAE exposure in elite American football and lay the foundation for future position-specific interventions aimed at reducing HAE burden and enhancing player safety.

**Methods**

A prospective observational cohort study was conducted using data from an NCAA Division 1 American Football team during the 2022 Mountain West Conference Regular Season. Thirty-five players were fitted with iMGs (Prevent Biometrics, Minneapolis, MN, USA). Participation was voluntary, and ethical approval was obtained from the University of Nevada Institutional Review Board (No. 1757959-5) and in accordance with the Declaration of Helsinki.

All iMGs were custom fit following three-dimensional dental scans with a dental assistant under the supervision of a certified dentist. The iMGs are equipped with triaxial gyroscopes and accelerometers sampling at 3200Hz with a measurement range of +/-35rad/s$^2$ and +/- 200g, respectively.[20] HAEs were identified when linear acceleration exceeded an 8g trigger threshold on any single axis of the iMG accelerometer. Linear acceleration was then transformed to the head centre of gravity. Time series data for HAEs were captured 10ms prior and 40ms after the trigger event, with a recording threshold of 5g of resultant peak linear acceleration (PLA) and resultant peak angular acceleration (PAA) of 400rad/s$^2$ at the head centre of gravity.[16,21] The degree of noise/artifact in each data signal was classified by a Prevent Biometrics proprietary algorithm as minimal (class 0), moderate (class 1), or severe (class 2). A 4-pole (2x2), zero-phase, low-pass Butterworth filter was applied to each kinematic signal, similar to previous studies.[19,21] A 200Hz, 100Hz and 50Hz cut-off frequency (-6dB) were applied to class 0, class 1, and class 2 signals, respectively.[20] Proximity sensor readings were collected to ensure that iMGs were seated to the player's dentition during HAEs.[19,21] NCAA footage was collected from mainstream television broadcasters, and all video analysis was performed by a trained video analyst with over 6 years of experience.[22–24]

The sample comprised of 35 players participating in a total of 71 player matches (games) across 4 games. The validation protocol was utilised from Jones et al.[12] All video analysis was conducted using QuickTime Player v.10.5 on a 32" monitor. The reviewer was permitted to pause, rewind, and zoom as needed. First, an unguided video analysis of all footage was performed with the reviewer blinded to all iMG data. The reviewer looked for instances of direct and visible contact to the head during gameplay, affecting any players actively wearing iMGs. The video time of each event was recorded in a list, organised by player and match. A total of 517 head impacts were identified in the unguided video analysis. Proximity sensor readings were used to exclude any recorded events where the iMG was 'off the teeth' (n=31). All events that did not meet the 5g and 400rad/s$^2$ threshold were also excluded (n=32). As such the sample size for unguided analysis was 454 impacts.

Next, the iMG data was synchronised with the video footage to ensure frame-to-frame synchronisation similar to Jones et al.[12] The iMG dataset was cross-referenced with the list of head impact events from the unguided video analysis. Each event with a corresponding iMG-triggered event was designated a True Positive and any event without was designated a False Negative. Sensitivity with a 95% confidence interval (CI) was calculated (Eq. 1).

$$Sensitivity = \frac{True\ Positives}{True\ Positives + False\ Negatives} \quad [1]$$

For the guided analysis, there were a total of 1101 iMG triggered events that met the 5g and 400rad/s$^2$ recording threshold. Each of these iMG triggered events was cross-referenced with the video footage and designated as either True Positive (TP), False Positive (FP), False Positive Off-Camera (FPOC) or False Positive Inactive (FPI) based on the definitions in Table 1.[12] Positive predictive values (PPVs) with 95% CI were calculated for TPs and FPs, both including and excluding FPOC and FPI events (Eq. 2).

$$PPV = \frac{True\ Positives}{True\ Positives + False\ Positives} \quad [2]$$

Only true positive iMG-triggered events (i.e., HAEs) that met the 5g and 400rad/s$^2$ threshold were used in the incidence calculations. Proximity sensor data confirmed that iMGs were being worn during each analysed event. Only player matches where the instrumented player wore the iMG for at least 90% of their contact events were used (n=64, 90.1%) in the incidence calculations.[19,21] The approach produced a total of 828 HAEs, and 64 eligible player matches over 4 games. Data for the incidence analysis consisted of individual HAEs recorded across Defense and Offense positions. Positions were consolidated for analysis. Offensive positions comprised Offensive Linemen (Centers, Guards, Long Snappers, and Offensive Tackles; n=6), Running Backs (n=2), and Tight Ends (n=4), while Defensive positions comprised of Defensive Linemen (Defensive Ends and Defensive Tackles; n=2), Linebackers (n=14), and Defensive Backs (Safeties and Cornerbacks; n=7).

**Table 1:** Definitions of True Positive (TP), False Positive (FP), False Positive Off-Camera (FPOC) or False Positive Inactive (FPI) for guided and unguided analysis.[12]

| Category | Guided Analysis | Unguided Analysis |
|---|---|---|
| True Positive (TP) | iMG-triggered event corresponds with a contact event, i.e. direct head contact or inertial head loading during active gameplay. | An instance of direct head contact witnessed on video footage in an instrumented player corresponds with a recorded HAE. |
| False Positive (FP) | iMG-triggered event does not correspond with a contact event on video footage during active gameplay. | Not Applicable |
| Assumed False Positive Off-Camera (FPOC) | iMG-triggered event occurs during active match play while the player is on the field. However, the player is not visible in video footage. This is assumed to be false-positive. | Not Applicable |
| Assumed False Positive Inactive (FPI) | iMG-triggered event occurs when active gameplay is not ongoing, there is a commercial break, or the instrumented player is on the sidelines. | Not Applicable |
| False-Negative | Not Applicable | An instance of direct head contact is witnessed on video footage in an instrumented player, but no corresponding HAE is recorded by iMG. |

Resultant peak linear acceleration (PLA) and resultant peak angular acceleration (PAA) were extracted from each HAE recorded. HAE Incidence was calculated as the number of HAEs per player match. Mean values for Offense and Defense were calculated across players and 95% CIs were estimated using a bootstrapping procedure.[19] The dataset was randomly resampled 2500 times and the 2.5th and 97.5th percentile of resampled means were used as the lower and upper bounds of CI, respectively.[19] Mean incidence along with 95% CIs, were calculated across a range of PLA and PAA magnitude thresholds for Offense and Defense and a significant difference was assumed if the CIs did not overlap.[19]

## Results

The total number of TP, FP, FPOC, FPI events with PPV and Sensitivity calculations are seen in Table 2. On-field validity of the Prevent Biometrics iMG in NCAA American Football indicates a sensitivity was 0.89 and PPV ranging from 0.76-0.98 based on false positive definitions (Table 2). The incidence of PLA and PAA HAEs above a range of thresholds in Defense and Offense appear similar (Figure 1). The incidence of HAEs above 10 g was 11.2 and 11.3 HAEs per player match for Defense and Offense, respectively, while PAA incidence above 1.0 krad/$s^2$ was 5.5 and 6.9 HAEs per player match for Defense and Offense, respectively (Fig. 1). Incidence of HAEs above 30 g was 1.6 and 2.6 per player match and 0.9 and 1.4 for HAEs above 2.0 krad/$s^2$ for Defense and Offense, respectively.

**Table 2**: On-field validity of the Prevent Biometrics iMG in NCAA Football at recording threshold of 5g and 400rad/$s^2$

| Unguided Analysis | Value |
| --- | --- |
| True Positive; TP (n) | 403 |
| False Negative; FP (n) | 51 |
| Sensitivity (95% CI) | 0.89 (0.86-0.92) |
| **Guided Analysis** | **Value** |
| True Positive; TP(n) | 828 |
| False Positive; FP (n) | 19 |
| False Positive Off Camera; FPOC (n) | 46 |
| False Positive Inactive; FPI (n) | 208 |
| All assumed false positives (n) | 254 |
| PPV: True Positive and False Positive (95% CI) | 0.98 (0.97-0.99) |
| PPV: True positive, false positive, false positive off camera (95% CI) | 0.93 (0.91-0.94) |
| PPV: True positive and all assumed false positive (95% CI) | 0.76 (0.73-0.78) |

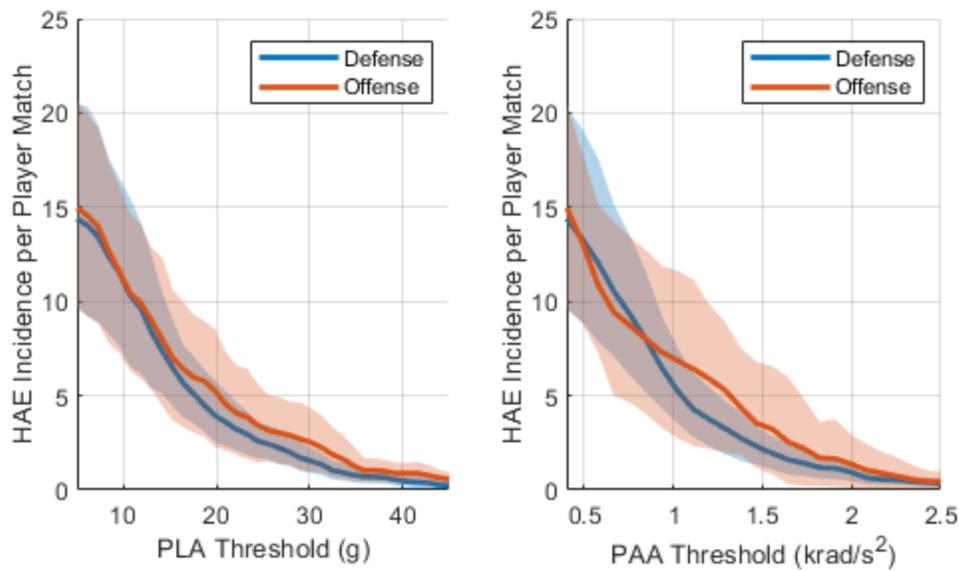

**Figure 1**: Incidence of HAEs for NCAA American Football players for Defense and Offense positional groups.

**Discussion**

Previous studies have validated the accuracy of the Prevent Biometrics iMG in both laboratory and on-field settings. In rugby, similar on-field video-verification validation studies using the Prevent iMG produced comparable results. Considering True Positives and False Positives, Jones et al.[12] reported a PPV of 0.94 (95% CI 0.92-0.95; no recording threshold applied) which was comparable to 0.98 (95% CI 0.97-0.99) in the current study. Tooby et al.[19] applied the same 5g and 400rad/s$^2$ recording threshold as in the present study and reported a PPV of 0.99 (95% CI 0.97-1.00). The sensitivity value of 0.89 (95% CI 0.86-0.92) in the current study resembled rugby data presented by Jones et al.[12] and Tooby et al.[19] who reported 0.75 and 0.86, respectively. The findings are also similar to the FRI-Biocore iMG system that achieved a PPV of 0.98 and sensitivity of 0.82 in American football.[15] Overall, our data supports the validity of the Prevent iMG in measuring in-game HAEs.

HAE incidence per player match provides an approximation of player exposure to contact events on a match-to-match basis. HAEs above 30g or 2.0krad/s$^2$, which have previously been considered higher magnitude,[19] were relatively rare. HAE incidence between Offense and Defense appeared similar. Offensive Linemen are typically some of the largest players on a team, predominantly responsible for protecting the Quarterback and creating running lanes.[25,26] This can results in frequent, lower-speed collisions at a short range. In contrast, defensive backs are marginally smaller, more agile players who tackle fast-moving players like the running back.[25,27] Somewhere in between these two roles is the Tight End, who can serve as a receiver, ball carrier, or blocker depending on the play.[25,28] This likely exposes them to a mix of lower and high-speed impacts.

Research on concussion incidence in elite American Football has identified Tight End, Running Back and Cornerbacks as the most diagnosed, followed by Defensive Ends and Linebackers.[29] Offensive Tackles, Guards, Centers and Long-Snappers reported some of the lowest rates. The use of iMGs in sport compliments these findings by providing information on HAE frequency and severity that players can be exposed to over a playing career allowing potential risk mitigation strategies to be proactively implemented.[30]

To reduce match HAE exposure in active players, approaches can aim to (1) reduce match exposure, (2) minimise the frequency of HAEs and (3) minimise the magnitude of HAEs. Optimising diagnostic techniques for both acute and chronic brain injury through protocol assessments and longitudinal cognitive evaluation will ideally include HAE exposure if it has the potential to be damaging. It is thought that susceptibility to concussion increases with increased HAE exposure.[31] Limiting full-contact reps in training[32] and emphasising safer tackling techniques, such as discouraging players from tackling with their helmet, would likely help reduce HAE exposure.[4,33,34] iMGs offer a means of measuring HAE incidence and exposure, but the threshold at which intervention might be required remains unclear. Data collected from iMGs might be used to retrospectively evaluate cumulative exposure faced by players who suffer a concussion, prior to their diagnosis. Future research might also involve larger sample sizes of iMG data and evaluate HAE mechanisms across more granular positional groups, which could be used to create tailored mitigation strategies for each position and explore the effect of contact technique on HAE incidence and propensity.

This study was of a limited sample size, spread unevenly across playing positions. Players switching positions temporarily, which happens occasionally in American Football, was not controlled for. The dataset was limited to HAEs recorded during match play and did not include training sessions where HAEs can occur.[32] The signal processing of noisy signals performed in-house by Prevent Biometrics and the proximity sensor data used to ensure rigid fit to the dentition have not been independently validated but used in similar studies.[19,30] The findings are limited to the Prevent Biometrics iMG system. Signal processing methods such as HEADSport have been developed to enable data comparison across different iMG systems.[35] This study used only peak linear and angular head kinematics, which do not account for directionality or temporality (i.e. pulse duration) which may influence injury risk and/or HAE severity.

**Conclusion**

iMGs were validated and subsequently used to quantify HAE incidence in elite NCAA American Football. The high sensitivity and PPV is comparable to un-helmeted sports such as Rugby, suggesting that the iMGs are suitable for use in American Football. Position-based variability in HAE incidence and magnitude highlights the need for tailored mitigation strategies, though offensive and defensive players showed similar overall rates. This study provides a benchmark assessment of HAE incidence in elite American Football. The data generated in this study lays a foundation for the development of position-specific interventions aimed at reducing HAE exposure. A more comprehensive understanding of the complex relationship between HAEs, concussion risk and long-term brain health is needed to pave the way for evidence-based strategies to enhance player safety and mitigate the risk of long-term neurological sequelae.

**REFERENCES**

1. Lessley DJ, Kent RW, Funk JR, et al. Video Analysis of Reported Concussion Events in the National Football League During the 2015-2016 and 2016-2017 Seasons. *Am J Sports Med*. 2018;46(14):3502-3510. doi:10.1177/0363546518804498

2. Zuckerman SL, Kerr ZY, Yengo-Kahn A, Wasserman E, Covassin T, Solomon GS. Epidemiology of Sports-Related Concussion in NCAA Athletes From 2009-2010 to 2013-2014: Incidence, Recurrence, and Mechanisms. *Am J Sports Med*. 2015;43(11):2654-2662. doi:10.1177/0363546515599634

3. Stern RA, Riley DO, Daneshvar DH, Nowinski CJ, Cantu RC, McKee AC. Long-term consequences of repetitive brain trauma: chronic traumatic encephalopathy. *PM R*. 2011;3(10 Suppl 2):S460-467. doi:10.1016/j.pmrj.2011.08.008

4. Pankow MP, Syrydiuk RA, Kolstad AT, et al. Head Games: A Systematic Review and Meta-analysis Examining Concussion and Head Impact Incidence Rates, Modifiable Risk Factors, and Prevention Strategies in Youth Tackle Football. *Sports Med*. 2022;52(6):1259-1272. doi:10.1007/s40279-021-01609-4

5. Tucker R, Raftery M, Kemp S, et al. Risk factors for head injury events in professional rugby union: a video analysis of 464 head injury events to inform proposed injury prevention strategies. *Br J Sports Med*. 2017;51(15):1152-1157. doi:10.1136/bjsports-2017-097895

6. Walton-Fisette T. Concussions in NCAA Football. *Journal of Sport History*. 2024;51(2):33-45. doi:10.5406/21558450.51.2.04

7. McKee AC, Stein TD, Nowinski CJ, et al. The spectrum of disease in chronic traumatic encephalopathy. *Brain*. 2013;136(1):43-64. doi:10.1093/brain/aws307

8. Mez J, Daneshvar DH, Kiernan PT, et al. Clinicopathological Evaluation of Chronic Traumatic Encephalopathy in Players of American Football. *JAMA*. 2017;318(4):360. doi:10.1001/jama.2017.8334

9. Mez J, Daneshvar DH, Abdolmohammadi B, et al. Duration of American Football Play and Chronic Traumatic Encephalopathy. *Annals of Neurology*. 2020;87(1):116-131. doi:10.1002/ana.25611

10. Tierney GJ, Kuo C, Wu L, Weaving D, Camarillo D. Analysis of head acceleration events in collegiate-level American football: A combination of qualitative video analysis and in-vivo head kinematic measurement. *Journal of Biomechanics*. 2020;110:109969. doi:10.1016/j.jbiomech.2020.109969

11. Wu LC, Nangia V, Bui K, et al. In Vivo Evaluation of Wearable Head Impact Sensors. *Annals of Biomedical Engineering*. Published online 2016. doi:10.1007/s10439-015-1423-3

12. Jones B, Tooby J, Weaving D, et al. Ready for impact? A validity and feasibility study of instrumented mouthguards (iMGs). *Br J Sports Med*. 2022;56(20):1171-1179. doi:10.1136/bjsports-2022-105523

13. Kieffer EE, Begonia MT, Tyson AM, Rowson S. A Two-Phased Approach to Quantifying Head Impact Sensor Accuracy: In-Laboratory and On-Field Assessments. *Ann Biomed Eng*. 2020;48(11):2613-2625. doi:10.1007/s10439-020-02647-1

14. Quigley KG, Hopfe D, Fenner M, et al. Preliminary Examination of Guardian Cap Head Impact Kinematics Using Instrumented Mouthguards. *J Athl Train*. 2024;59(6):594-599. doi:10.4085/1062-6050-0136.23

15. Gabler LF, Huddleston SH, Dau NZ, et al. On-Field Performance of an Instrumented Mouthguard for Detecting Head Impacts in American Football. *Ann Biomed Eng*. 2020;48(11):2599-2612. doi:10.1007/s10439-020-02654-2

16. Tooby J, Weaving D, Al-Dawoud M, Tierney G. Quantification of Head Acceleration Events in Rugby League: An Instrumented Mouthguard and Video Analysis Pilot Study. *Sensors*. 2022;22(2):584. doi:10.3390/s22020584

17. Sokol-Randell D, Stelzer-Hiller OW, Allan D, Tierney G. Heads Up! A Biomechanical Pilot Investigation of Soccer Heading Using Instrumented Mouthguards (iMGs). *Applied Sciences*. 2023;13(4):2639. doi:10.3390/app13042639

18. Liu Y, Domel AG, Yousefsani SA, et al. Validation and Comparison of Instrumented Mouthguards for Measuring Head Kinematics and Assessing Brain Deformation in Football Impacts. *Ann Biomed Eng*. 2020;48(11):2580-2598. doi:10.1007/s10439-020-02629-3

19. Tooby J, Woodward J, Tucker R, et al. Instrumented Mouthguards in Elite-Level Men's and Women's Rugby Union: The Incidence and Propensity of Head Acceleration Events in Matches. *Sports Med*. 2024;54(5):1327-1338. doi:10.1007/s40279-023-01953-7

20. Woodward J, Tooby J, Tucker R, et al. Instrumented mouthguards in elite-level men's and women's rugby union: characterising tackle-based head acceleration events. *BMJ Open Sport Exerc Med*. 2024;10(3):e002013. doi:10.1136/bmjsem-2024-002013

21. Allan D, Tooby J, Starling L, et al. The Incidence and Propensity of Head Acceleration Events in a Season of Men's and Women's English Elite-Level Club Rugby Union Matches. *Sports Med*. 2024;54(10):2685-2696. doi:10.1007/s40279-024-02064-7

22. Armstrong N, Rotundo M, Aubrey J, Tarzi C, Cusimano MD. Characteristics of potential concussive events in three elite football tournaments. *Injury Prevention*. Published online 2019. doi:10.1136/injuryprev-2019-043242

23. Sokol-Randell D, Rotundo, MP, Denley C, et al. Characteristics and Assessment of Potential Concussive Events in the UEFA Champions League. In: Brain Injury; 2021.

24. Rotundo MP, Sokol-Randell D, Bleakley C, Donnelly P, Tierney G. Characteristics of potential concussive events in elite hurling: a video-analysis study. *Ir J Med Sci*. 2023;192(6):3175-3185. doi:10.1007/s11845-023-03307-8

25. Pincivero DM, Bompa TO. A Physiological Review of American Football: *Sports Medicine*. 1997;23(4):247-260. doi:10.2165/00007256-199723040-00004

26. Coach Martin. Offensive Line Positions in Football (All Roles Explained). Football Advantage. February 9, 2025. https://footballadvantage.com/offensive-line-positions/

27. Coach Martin. What is a Linebacker in Football? (LB Position Guide). Football Advantage. February 9, 2025. https://footballadvantage.com/linebacker/

28. Coach Martin. What is a Tight End in Football? (TE Position Guide). Football Advantage. February 9, 2025. https://footballadvantage.com/tight-end/

29. Dai JB, Li AY, Haider SF, et al. Effects of Game Characteristics and Player Positions on Concussion Incidence and Severity in Professional Football. *Orthopaedic Journal of Sports Medicine*. 2018;6(12):2325967118815448. doi:10.1177/2325967118815448

30. Allan D, Tooby J, Starling L, et al. Head Kinematics Associated with Off-Field Head Injury Assessment (HIA1) Events in a Season of English Elite-Level Club Men's and Women's Rugby Union Matches. *Sports Med*. Published online November 16, 2024. doi:10.1007/s40279-024-02146-6

31. Abrahams S, Fie SM, Patricios J, Posthumus M, September AV. Risk factors for sports concussion: an evidence-based systematic review. *Br J Sports Med*. 2014;48(2):91-97. doi:10.1136/bjsports-2013-092734

32. McCrea MA, Shah A, Duma S, et al. Opportunities for Prevention of Concussion and Repetitive Head Impact Exposure in College Football Players: A Concussion Assessment, Research, and Education (CARE) Consortium Study. *JAMA Neurol*. 2021;78(3):346. doi:10.1001/jamaneurol.2020.5193

33. Cross MJ, Tucker R, Raftery M, et al. Tackling concussion in professional rugby union: a case-control study of tackle-based risk factors and recommendations for primary prevention. *Br J Sports Med*. 2019;53(16):1021-1025. doi:10.1136/bjsports-2017-097912

34. Stockwell DW, Blalock R, Podell K, Marco RAW. At-Risk Tackling Techniques in American Football. *Orthopaedic Journal of Sports Medicine*. 2020;8(2):2325967120902714. doi:10.1177/2325967120902714

35. Tierney G, Rowson S, Gellner R, et al. Head Exposure to Acceleration Database in Sport (HEADSport): a kinematic signal processing method to enable instrumented mouthguard (iMG) field-based inter-study comparisons. *BMJ Open Sport Exerc Med*. 2024;10(1):e001758. doi:10.1136/bmjsem-2023-001758